\documentclass{aastex}
\usepackage{emulateapj5}
\usepackage{epsf,graphics}
\begin{document}
\title{Reliability of the dark matter clustering in cosmological
$N$-body simulation  on scales below the mean separation length of
particles}

\author{Takashi Hamana\altaffilmark{1,2}, Naoki 
Yoshida\altaffilmark{2,3}, and Yasushi Suto\altaffilmark{4}}
\affil{$^1$National Astronomical Observatory, Mitaka 181-8588, Japan\\
$^2$Max-Planck-Institut f\"ur Astrophysik,
Karl-Schwarzschild-Strasse 1,  85748 Garching, Germany\\
$^3$Harvard-Smithsonian Center for Astrophysics
60 Garden Street, Cambridge MA02138\\
$^4$Department of Physics and Research Center for the Early Universe
(RESCEU) \\  School of Science, University of Tokyo, Tokyo 113-0033,
Japan}
\email{hamana@yukawa.kyoto-u.ac.jp, ~ nyoshida@cfa.harvard.edu, ~
suto@phys.s.u-tokyo.ac.jp}

\received{2001 October 30}
\accepted{2002 December 4}

\begin{abstract}

We critically examine the reliability of the dark matter clustering in
high-resolution cosmological $N$-body simulations on scales below the
mean separation length of particles.  The particle discreteness effect
imposes the two fundamental limitations on those scales; the lack of the
initial fluctuation power and the finite mass resolution.  We address
this problem applying the dark halo approach and are able to discuss
separately how those two limitations affect the dark matter clustering
in $N$-body simulations at early epochs.  We find that limitations of
the dark matter clustering are primarily determined by the mass of
particles.  By a detailed comparison with three major cosmological
simulations, we also find that in order to reproduce a proper amplitude
of the dark matter clustering on small scales, halos with a
characteristic nonlinear mass, $M_{\rm NL}(z)$ defined by
$\sigma_R(M_{\rm NL}; z)=1$, must be resolved in the simulation.  This
leads to a critical redshift $z_{\rm crit}$ determined by $M_{\rm
NL}(z_{\rm crit})=n_{\rm halo} m_{\rm part}$ where $n_{\rm halo}$ is the
number of particles necessary to resolve the typical nonlinear mass halo
($\sim 10$).  We conclude that, at least as far as the two-point
correlation functions are concerned, the dark matter clustering in
high-resolution $N$-body simulations on scales below the mean particle
separation is reliable down to the gravitational force resolution length
only for $z<z_{\rm crit}$, while it is strongly affected by the finite mass
resolution for $z>z_{\rm crit}$.
\end{abstract}

\section{Introduction}

$N$-body simulations have played a key role in studying large-scale
structure formation in the universe.  In particular, they enable one to
probe the quasi-linear to nonlinear clustering in a direct manner
without resorting to any unrealistic approximations.  Thus they are
extensively used in cosmology as a powerful and indispensable tool
complementary to analytical techniques.  Cosmological $N$-body
simulations have been first applied in exploring the nonlinear growth of
two-point correlation functions (Miyoshi \& Kihara 1975; Aarseth, Gott,
\& Turner 1979), and in studying the halo mass function (Press \&
Schechter 1974). A seminal paper by Davis et al. (1985) clearly
demonstrated that this methodology provides a variety of testable
predictions for a given cosmological model.  Since then almost all
precise predictions in gravitational nonlinear phenomena have relied
heavily on N-body simulations, including accurate nonlinear modeling of
the dark matter power spectrum (Hamilton et al. 1991; Peacock \& Dodds
1996), redshift-space distortion (Suto \& Suginohara 1991; Cole, Fisher
\& Weinberg 1994; Magira, Jing \& Suto 2000), three- and four-point
correlation functions (Suto 1993; Matsubara \& Suto 1994; Suto \&
Matsubara 1994), probability distribution function of the nonlinear
density field (Suto, Itoh \& Inagaki 1990; Lahav et al. 1993; Kayo,
Taruya \& Suto 2001), the mass function of virialized dark halos
(Bahcall \& Cen 1993; Sheth \& Tormen 1999; Jenkins et al.~2001), the
density profiles of dark matter halos (Navarro, Frenk \& White
1996,1997; Fukushige \& Makino 1997; Moore et al. 1998; Jing \& Suto
2000; Yoshida et al. 2000), dark halo biasing (Watanabe, Matsubara, \&
Suto 1994; Jing \& Suto 1998; Jing 1998; Sheth \& Tormen 1999; Colberg
et al. 2000; Taruya et al. 2001), and clustering on the past light-cone
(Hamana et al.~2001a,b; Evrard et al.~2001).

Of course, the reliability of $N$-body simulations is severely limited
by various numerical restrictions. Among them, the effective dynamical
range is quite important. 
Consider a simulation employing $N_{\rm part}$
particles in a comoving cubic box of $L_{\rm box}^3$. Obviously the {\it
mass resolution} is set by the number of particles, and the
corresponding resolution in length scale is up to the mean particle
separation $\lambda_{\rm part} = L_{\rm box}/N_{\rm part}^{1/3}$. In
fact, the initial perturbations of simulations are usually set on
uniform grid points of an interval $\lambda_{\rm part}$.
Therefore one cannot properly set the desired initial power spectrum
for $k>k_{\rm Nyquist}$ where $k_{\rm Nyquist} =\pi/\lambda_{\rm
part}$ is the  Nyquist wavenumber\footnote{
This is also the case for a $N$-body simulation that adopts a
``glass'' distribution for particles to represent an initial uniform
state (White 1996, Jenkins et al.~1998), because the Nyquist
wavenumber in this case is also set by
$k_{\rm Nyquist} =\pi/\lambda_{\rm part}$.}.
Nevertheless most N-body codes adopt a {\it force
resolution} (e.g., the gravitational softening length $\epsilon_{\rm
grav}$) typically one order of magnitude smaller than $\lambda_{\rm
part}$.  Are these {\it high-resolution simulations} \footnote{Throughout the
present paper, we use the term ``high-resolution simulation'' to refer
to the $N$-body codes that employ the gravitational softening length
smaller than the mean particle separation such as P$^3$M and Tree
codes.} really reliable on scales below the particle mean separation
length $\lambda_{\rm part}$ where the initial conditions cannot be set
properly ? Although several attempts have been made to this direction (e.g.,
Efstathiou et al.~1985; 1988; Gelb \& Bertschinger 1994; Melott et
al.~1997; Splinter et al.~1998), 
this natural question has remained unanswered in a 
quantitative manner, and is exactly what
we address in detail in the rest of the paper.

Qualitatively speaking, high-resolution simulations are justified {\it
only if almost all the power on scales below the mean separation 
$\lambda_{\rm part}$ at later epochs is generated via nonlinear
mode-coupling from the power on larger scales}. The transfer of the
fluctuation power in cosmology due to the nonlinear mode-coupling is
examined numerically by Suginohara et al. (1991), as well as by
perturbation analysis in Suto \& Sasaki (1991) and Makino, Sasaki \&
Suto (1992). According to their results, the nonlinear power transfer is
expected to {\it eventually} dominate the initial power if the effective
spectral index at the scale, $n_{\rm eff} \equiv d \log P(k)/ d \log k$,
is much less than $-1$.  Actually this is indeed the case for most
realistic cosmological scenarios including cold dark matter (CDM) models
whose power spectrum asymptotically approaches $\propto k^{-3}$.
Nevertheless the above qualitative argument does not
answer precisely {\it when} the results below $\lambda_{\rm part}$ become
reliable.

To be more specific, we examine the evolution of dark matter clustering
in high-resolution cosmological $N$-body simulations taking explicit
account of the two fundamental limitations: (i) the lack of the initial
fluctuation power on scales smaller than the mean particle separation,
and (ii) the finite mass resolution.  To do this, we apply the dark halo
approach in which the nonlinear dark matter two-point correlation
function is described by a sum of two contributions; one from the
correlations between the halos (halo-halo correlation), and the other
from correlations of matter within the same halo (the Poisson
contribution) (Peebles 1980 and references therein, for recent
developments in the dark halo approach, see Ma \& Fry 2000; Seljak
2000).  We show in the following sections that this approach provides a
simple but powerful way to examine separately how those two limitations
affect the simulation results.  This approach is quite different from
previous ones which usually attempted to find convergence among
different N-body codes with different numerical resolutions (e.g.,
Efstathiou et al.~1985, 1988; Gelb \& Bertschinger 1994; Melott et
al. 1997; Splinter et al. 1998; Moore et al. 1998; Jenkins et al.~1998,
2001; Knebe et al. 2000).

We focus on the two-point correlation function of dark matter to examine
the limitations in high-resolution $N$-body simulations for the
following reasons. Firstly, the two-point correlation function is one of
the most fundamental statistics to quantify clustering properties, and
has been studied in detail both analytically and numerically.  Actually,
there exist accurate fitting formulae of the dark matter power spectrum
(Hamilton et al. 1991; Peacock \& Dodds 1996) that reproduce the
behavior of the two-point correlation function in linear to highly
nonlinear regimes.  Secondly, the estimator of the two-point correlation
function from the particle distribution in $N$-body data is well
established (Kerscher, Szapudi \& Szalay 2000 and references therein).
These two advantages enable us to investigate limitations in
high-resolution $N$-body simulations through detailed comparison between
numerical results and model predictions.
 
The outline of this paper is as follows.  In \S 2, we summarize our dark
halo approach, and demonstrate the extent to which the two limitations
may affect the evolution of dark matter clustering on scales below the
mean particle separation length. We present detailed comparison between
a high-resolution $N$-body data and our model predictions in \S 3, and
show that our model reproduces well an apparent depression in the dark
matter two-point correlation function measured from those simulation
data.  In particular we find that the depression becomes less
appreciable as the ratio of the characteristic nonlinear mass $M_{\rm
NL}$ to the particle mass $m_{\rm part}$ increases.
On the basis of this, we evaluate the critical
redshift $z_{\rm crit}$ beyond which the correlation function of the
dark matter in high-resolution $N$-body simulations on scales below
the mean particle separation length suffers significantly from
discreteness limitations.
Finally \S 4 is devoted to a summary and further discussion.

\section{Dark halo approach}

The dark halo approach has a long history (Neyman \& Scott 1952; Limber
1953; Peebles 1974, 1980; McClelland \& Silk 1977), and has recently
applied to various problems in galaxy formation and cosmological
nonlinear clustering (Scherrer \& Bertschinger 1991; White \& Frenk
1991; Kauffmann, White \& Guiderdoni 1993; Cole et al.~1994; Sheth \&
Jain 1997; Komatsu \& Kitayama 1999; Somerville \& Primack 1999; Seljak
2000; Peacock \& Smith 2000; Ma \& Fry 2000; Cooray, Hu \&
Miralda-Escude 2000; Cooray \& Hu 2001).  In this section, we summarize
several expressions which are most relevant to the current analysis.
Further details of this approach can be found in those papers.

\subsection{Ingredients in model predictions}

The accurate description of the matter clustering on the basis of the
dark halo approach requires three major ingredients; the mass
function of dark halos, the mass-dependent biasing factor of
halos, and the density profile of halos. We present a summary of those
three ingredients below.

We adopt an analytical model of the halo mass function proposed by Sheth \&
Tormen (1999):
\begin{eqnarray}
\label{eq:STmassfunc}
{{dn}\over{dM}}dM &=& {{\bar{\rho_0}} \over {M}} f(\nu)d \nu 
\nonumber \\
&=& {{\bar{\rho_0}} \over {M}} 
A [1+{(a\nu)}^{-p}]\sqrt{a\nu}\exp\left(-{ a\nu\over 2}\right)
\frac{d\nu}{\nu}, 
\end{eqnarray}
where
\begin{eqnarray}
\label{eq:nu-M}
\nu=\left[{{\delta_c(z)} \over {\sigma(M,z)}} \right]^2 ,
\end{eqnarray}
$\bar{\rho_0}$ is the mean cosmic mass density, and the numerical
coefficients $a$ and $p$ are empirically fitted from N-body simulations
as $a=0.707$ and $p=0.3$.  Here $\sigma(M,z)$ is the root-mean-square
fluctuations in the matter density top-hat smoothed over a scale $R_M
\equiv (3M/4\pi \bar{\rho_0})^{1/3}$ and $\delta_c(z)$ is the threshold
over-density for spherical collapse (see Nakamura \& Suto 1997 and Henry
2000 for useful fitting functions).  The normalization constant $A$ in
equation  ~(\ref{eq:STmassfunc}) is determined by requiring that
\begin{eqnarray}
{1\over {\bar{\rho_0}}} \int {{dn}\over {dM}} M dM = \int f(\nu)d \nu=1.
\end{eqnarray}

A simple analytic model of the halo bias was proposed by Mo \& White
(1996) applying the extended Press-Schechter theory (e.g., Lacey \& Cole
1993).  Jing (1998) and Sheth \& Tormen (1999) discussed the correction
for the mass-dependence, and the stochastic nature of the halo biasing
is considered subsequently by Taruya \& Suto (2000) and Somerville et
al.~(2001). For the present
purpose, we use the scale-independent model by Sheth \& Tormen (1999)
which reasonably reproduces $N$-body data:
\begin{equation}
b(\nu)=1+{{a\nu-1}\over{\delta_c}} +{{2p} \over {\delta_c (1+a\nu)}}.
\end{equation}
Finally we have to specify the density profile of dark matter halos.  We
adopt the NFW profile (Navarro, Frenk \& White 1996):
\begin{eqnarray}
\rho(r)={ \rho_s \over (r/r_s)(1+r/r_s)^2},
\end{eqnarray}
where $\rho_s$ and $r_s$ denote the scaling density and radius.  While
more recent simulations (Fukushige \& Makino 1997; Moore et al. 1998)
suggest that the inner profile is indeed steeper, we adopt the original
profile just for simplicity. It is conventional to introduce the
concentration parameter $c=r_{\rm vir}/r_s$, where $r_{\rm vir}$ is the
virial radius of the halo. In terms of this, the mass of halo is written
as
\begin{eqnarray}
\label{eq:massc}
M=\frac{4\pi \rho_s r_{\rm vir}^3}{c^3}
\left[ \log(1+c)-{c \over 1+c} \right] .
\end{eqnarray}
Since the spherical collapse model indicates that $M=4 \pi r_{\rm vir}^3
\delta_{\rm vir}(z) \bar{\rho_0}/3$ with $\delta_{\rm vir}(z)$ being the
over-density of collapse (see, Nakamura \& Suto 1997; Henry 2000), one
can express $\rho_s$ in terms of $M$ and $c$ using equation
(\ref{eq:massc}).  Thus our description of dark halos is completed if
the concentration parameter $c$ is expressed as a function of $M$ and
$z$.  We adopt the following empirical relations:
\begin{equation}
\label{eq:c-M}
c(M,z) = \cases{\displaystyle
\frac{8}{1+z} \left(\frac{M}{10^{14}h^{-1}M_{\odot}}\right)^{-0.13}
& for $\Lambda$CDM,\cr
\displaystyle
\frac{5}{1+z} \left(\frac{M}{10^{14}h^{-1}M_{\odot}}\right)^{-0.13}
& for SCDM.\cr}
\end{equation}
The above $c$-$M$ relation for $\Lambda$CDM model is suggested by
Bullock et al.~(2001), while for SCDM model, we adopt the smaller
normalization found by Bartelmann et al. (1998).  For simplicity, we do
not take into account the scatter in $c$-$M$ relation (Bullock et
al.~2001), but it should have only secondary importance on a statistical
measurement such like the two-point correlation function of the dark
matter.  We compare the dark matter power spectrum predicted by the dark
halo approach (see next subsection) with that computed by the nonlinear
fitting formula of Peacock \& Dodds (1996). We find a good agreement
between them for the SCDM and $\Lambda$CDM models over the range
$0.01<k$[$h$Mpc$^{-1}]<100$ and over $0<z<5$.

\vspace{0.3cm}
\centerline{{\vbox{\epsfxsize=8.7cm\epsfbox{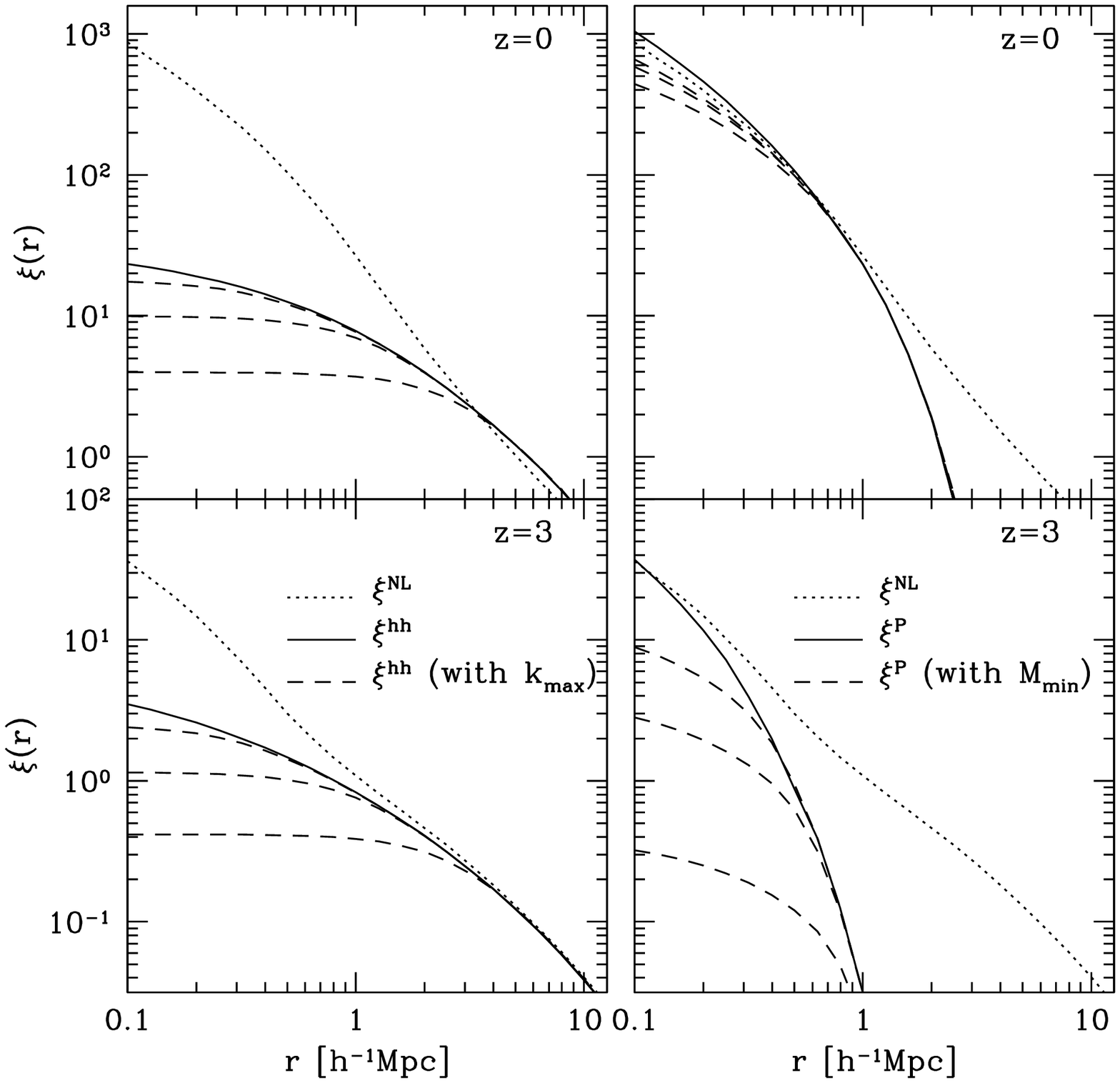}}}} 
\figcaption{Comparison between dark matter two-point correlation
functions predicted from the dark halo approach with and
without taking into account the two limitations in the $N$-body simulation.
{\it Left panels:} The effect of the lack of the initial perturbation
power on small scales is demonstrated. 
The dotted line shows the nonlinear correlation function computed using 
Peacock \& Dodds (1996) fitting function.
The solid line shows the two-point correlation function from the halo-halo
contribution without taking into account the small-scale cutoff, while the
dashed lines from upper to lower are for
$R_{\rm min}=2\pi/k_{\rm max}=0.3$, 1, 3$h^{-1}$Mpc. 
Top and bottom panels are for $z=0$ and $z=3$, respectively.
{\it Right panels:} The effect of the finite mass resolution is
demonstrated. The solid line shows the correlation function from
the Poisson contribution in the case of no low mass cutoff, while the
dashed lines from upper to lower are for $M_{\rm min}=1$,
$3$, 10$\times 10^{11} h^{-1}M_\odot$. \label{fig:halo_approach}}
\vspace{0.3cm}

\subsection{Predicting two-point correlation functions}

In the dark halo approach, the two-point correlation function of
the dark matter is the sum of two contributions; $\xi(r)=\xi^{\rm
hh}(r)+\xi^{\rm P}(r)$.  The first term (``the halo-halo contribution'')
arises due to the correlation between two halos of density profile
$\rho(r)$ that are biased tracers of the dark matter distribution.  In
Fourier space, this term can be explicitly written as (Seljak 2000):
\begin{eqnarray}
\label{eq:Phh}
P^{\rm hh}(k)=P_{\rm lin}(k) \left[ \int d\nu\, f(\nu) b(\nu)
y(k;M)\right]^2,
\end{eqnarray}
where $P_{\rm lin}(k)$ is the linear power spectrum, $y(k;M)$ is the
Fourier transform of the halo profile normalized by its mass,
$y(k;M)=\tilde{\rho}(k,M)/M$, and $M$ is related to $\nu$ via equation
~(\ref{eq:nu-M}).  The second ``Poisson term'' is due to the correlation
between dark matter particles in the same halo and is expressed as
\begin{eqnarray}
\label{eq:Pp}
P^{\rm P}(k)=\int d\nu\,f(\nu) {{M(\nu)}\over {\bar{\rho_0}}} 
\vert y(k;M) \vert^2.
\end{eqnarray}
Then the dark matter two-point correlation function is simply given by
the sum of two terms
\begin{eqnarray}
\xi^X(r) &=& {1\over {2\pi^2}} \int d k\,k^2 \left[ {{\sin(kr)}
\over {kr}} \right] P^X(k),
\end{eqnarray}
where $X$ stands for either hh or P for the halo-halo and Poisson terms,
respectively.

\vspace{0.3cm}
\centerline{{\vbox{\epsfxsize=8.7cm\epsfbox{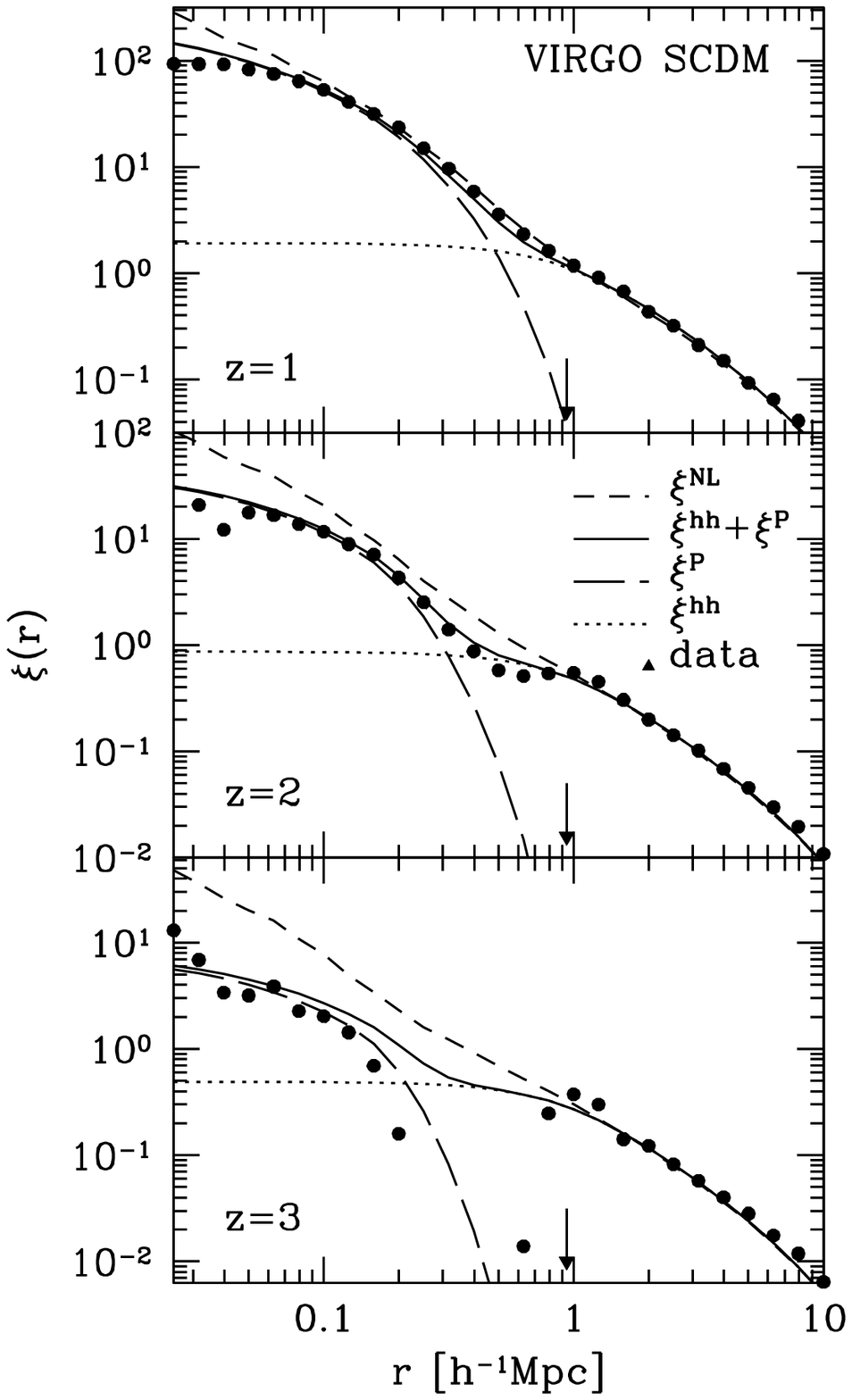}}}}
\figcaption{The two-point correlation function measured from the Virgo
SCDM $N$-body simulation (filled circles) are compared with predictions.
Dashed lines shows the nonlinear prediction by Peacock \& Dodds (1996).
Predictions by the dark halo approach, to which corrections for
the lack of small-scale perturbation power and the finite particle mass
are made (see \S 2), are shown by dotted, long-dashed and solid lines
for the halo-halo term, Poisson term and their sum, respectively.  The
arrows indicate the mean separation lengths $\lambda_{\rm part}$ of
simulation particles.  The redshift of the realization is indicated in
each panel. \label{fig:virgo_scdm}}
\vspace{0.3cm}

\subsection{Effects of the lack of small-scale power and the finite
mass resolution \label{subsec:limitations}}

Before presenting the detailed comparison between simulations and our
model predictions, it is instructive to show the main features that we
expect in the present methodology.  Figure \ref{fig:halo_approach} plots
the dark matter two-point correlation functions at $z=0$ ({\it Upper
panels}) and $z=3$ ({\it Lower panels}) for the set of parameters
employed for the Virgo $\Lambda$CDM simulation (see Table
\ref{table:models}).  Solid lines in the left and right panels show the
halo-halo and the Poisson terms, respectively.  For comparison, dotted
lines indicate the full nonlinear two-point correlation functions of the
dark matter computed with the Peacock \& Dodds (1996) formula. Clearly
the large-scale power is well described by the halo-halo contribution
only, while the small-scale power is dominated by the Poisson
contribution.

\begin{table*}
\caption{Parameters adopted in the simulation models.}
\label{table:models}
\begin{tabular}{lcccccccccc}
\hline
{} & $\Omega_{\rm m}$ & $\Omega_\Lambda$ & $\Omega_{\rm b}$ & $h$ 
& $\sigma_8$ & $N_{\rm part}$ & $z_{\rm init}$ 
& $m_{\rm part}$ & $\lambda_{\rm part}$ & $ \epsilon_{\rm grav}$\\
{} & & & & & & & & ($h^{-1}M_\odot$) & ($h^{-1}$Mpc) & ($h^{-1}$Mpc)\\ 
\hline
VIRGO-SCDM & 1.0 & 0.0 & 0.0 & 0.5 & 0.51 & 
$256^3$ & 50 & $2.27\times 10^{11}$ & 0.94 & 0.03\\
VIRGO-$\Lambda$CDM & 0.3 & 0.7 & 0.0 & 0.7 & 0.9 &  
$256^3$ & 30 & $6.86\times 10^{10}$ & 0.94 & 0.03\\
HUBBLE-$\Lambda$CDM & 0.3 & 0.7 & 0.04 & 0.7 & 0.9 & 
$10^9$ & 35 & $2.22\times 10^{12}$ & 3.0 & 0.1\\ 
\hline
\end{tabular}
\end{table*}

To simulate the effect of the numerical limitations more quantitatively,
we artificially set
\begin{eqnarray}
\label{eq:Phh_cut}
P^{\rm hh}(k)= 0 ~~{\rm for}~~k > k_{\rm max}
\end{eqnarray}
for the halo-halo term, and 
\begin{eqnarray}
\label{eq:STmassfunc_cut}
{dn\over dM} = 0  ~~{\rm for}~~M < M_{\rm min} 
\end{eqnarray}
for the Poisson term. The latter is equivalent to integrating equation
~(\ref{eq:Pp}) only for $\nu > \nu_{\rm min} \equiv
[\delta_c(z)/\sigma(M_{\rm min},z)]^2$.  The condition
(\ref{eq:Phh_cut}) corresponds to the lack of the initial fluctuation
power beyond the Nyquist wavenumber $k_{\rm Nyquist} \sim k_{\rm max}$,
while the condition (\ref{eq:STmassfunc_cut}) reflects the finite mass
resolution corresponding to the simulation particles.  In principle, the
condition (\ref{eq:STmassfunc_cut}) also affects the halo-halo term,
equation ~(\ref{eq:Phh}), however we do not include it, because our aim
here is to extract the most essential feature considering the most
important defect in the numerical simulation for each term.  We have
found, however, that including the correction results in only a minor
change in the estimates for the $P^{hh}$ term, so we do not consider the
condition (\ref{eq:STmassfunc_cut}) in the halo-halo term below.
Specifically, the dashed lines plotted in left and right panels of
Figure \ref{fig:halo_approach} adopt $R_{\rm min} \equiv \pi/k_{\rm
max} =0.3$, 1, 3$h^{-1}$Mpc, and $M_{\rm min}= 10^{11} h^{-1}M_\odot$,
$3\times 10^{11} h^{-1}M_\odot$, $10^{12} h^{-1}M_\odot$, from top to
bottom.

The message from these plots is that the dark matter clustering below
the mean separation length is really sensitive to the two parameters
$k_{\rm max}$ and $M_{\rm min}$.  In particular, the effect is quite
substantial at higher redshifts where the major fraction of mass is
contained in less massive halos that cannot be resolved with a given
particle mass in simulations.  In the next section we compare our model
predictions with the evolution of simulation data.

\section{Comparison with $N$-body data}

\subsection{N-body data}

For the present purpose, we need the current best simulation data.  So
we decided to use VIRGO-SCDM and VIRGO-$\Lambda$CDM simulations (Jenkins
et al.~1998) and HUBBLE-$\Lambda$CDM (Jenkins et al. 2001; Evrard et
al.~2001), all of which
are publicly available. The simulation parameters are summarized in
Table \ref{table:models}.  Their initial conditions were generated using
the CDM transfer functions computed by Bond \& Efstathiou (1984) for
the VIRGO simulations, while CMBFAST (Seljak \& Zaldarriaga 1996)
is used for the Hubble volume simulation which is well approximated by the
analytic fitting function by Eisenstein \& Hu (1998).
For the VIRGO simulations, we use snapshot
data at $z=1$, 2 and 3 for SCDM, and at $z=1$, 3 and 5 for $\Lambda$CDM.
In the case of the Hubble volume $\Lambda$CDM simulation, we do not use
the full snapshot data at $z$, but rather ``deep wedge'' light-cone
output. This dataset subtends an 81.45 square-degree field from its apex (a
fiducial observer position) at a corner of the simulation box directed
along a diagonal of the simulation box up to $z=4.9$ (Hamana et
al. 2001b)\footnote{For details of the output formats of the Hubble
volume simulation, see the Hubble volume project home page:
http://www.mpa-garching.mpg.de/Virgo/hubble.html}.  From this light-cone
output, we generate three slice subsamples with depth of $200h^{-1}$Mpc
centered at $z=0.5$, 1.5 and 2.5.

We compute the two-point correlation function of dark matter particles
using the pair-count estimator proposed by Landy \& Szalay (1993):
\begin{equation}
 \xi(r)= \frac{DD(r)-2DR(r)+RR(r)}{RR(r)}.
\end{equation}
In so doing, we distribute the same number of random particles as
simulation particles for each model.

\vspace{0.3cm}
\centerline{{\vbox{\epsfxsize=8.7cm\epsfbox{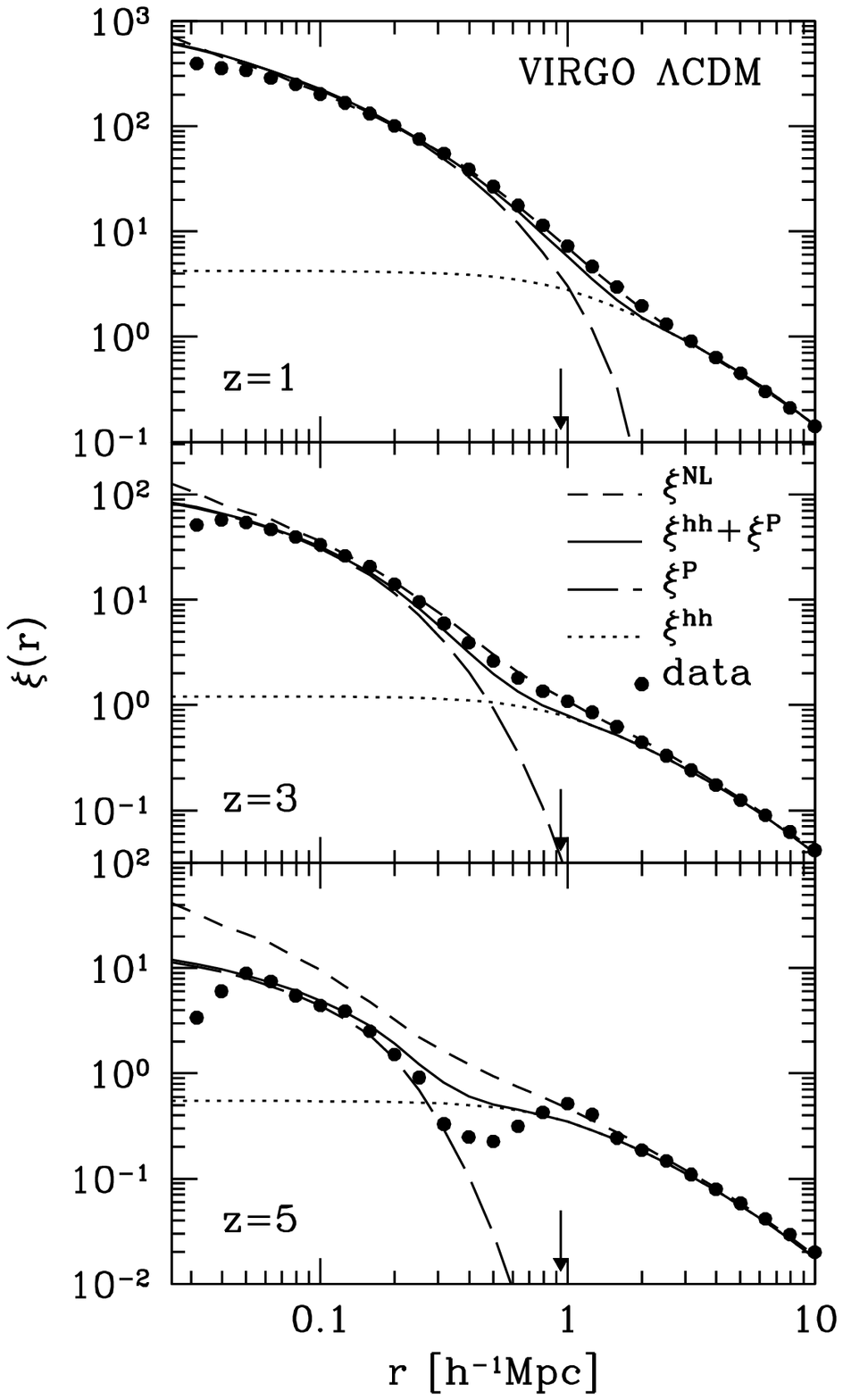}}}}
\figcaption{Same as Fig.~\ref{fig:virgo_scdm} but for VIRGO-$\Lambda$CDM
simulation.\label{fig:virgo_lcdm} }

\centerline{{\vbox{\epsfxsize=8.7cm\epsfbox{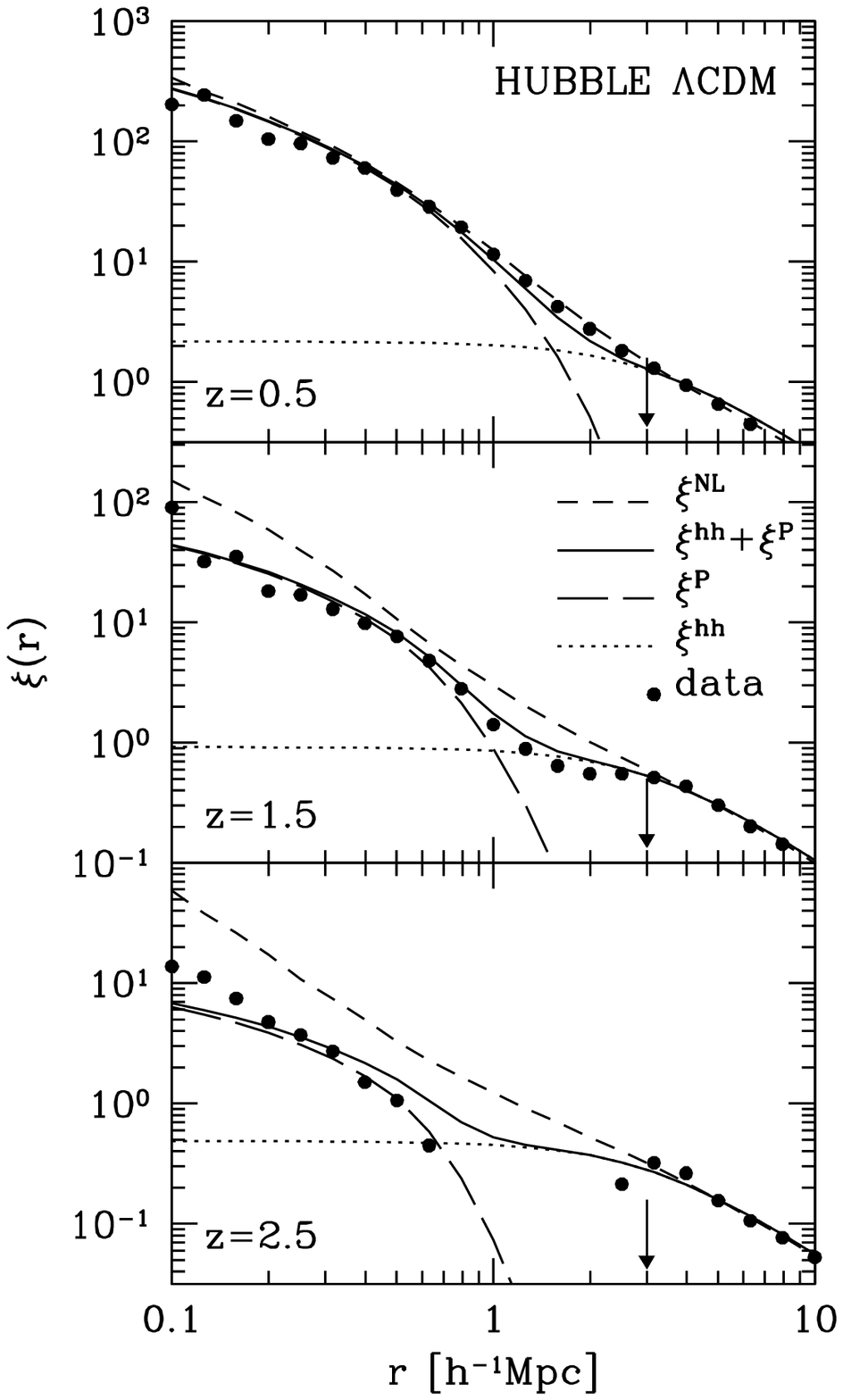}}}}
\figcaption{Same as Fig.~\ref{fig:virgo_scdm} but for Hubble volume
$\Lambda$CDM simulation. \label{fig:hubble_lcdm}}
\vspace{0.3cm}

\subsection{Model predictions}

We compute model predictions on the basis of the dark halo
approach which take into account the limitations due to the lack of the
initial fluctuation power and the finite mass resolution by introducing
the two parameters $k_{\rm max}$ and $M_{\rm min}$ as described in \S
\ref{subsec:limitations}. For definiteness we set $k_{\rm max}$ as
$k_{\rm Nyquist}= \pi/\lambda_{\rm part}$ for each simulation. The
minimum mass scale whose density profile can be resolved in simulations
is determined by the particle mass $m_{\rm part}$, but the precise value
is somewhat uncertain. In the analysis below we use $M_{\rm min}=\mu
m_{\rm part}$ with $\mu=4$, which, we find, reasonably fits the $N$-body
data almost independently of either the redshift or simulation
parameters.

\subsection{Evolution of the dark matter clustering at $r<
\lambda_{\rm part}$}

Figures \ref{fig:virgo_scdm} to \ref{fig:hubble_lcdm} present the
comparison between the simulation data and our model predictions for
Virgo SCDM, Virgo $\Lambda$CDM and Hubble $\Lambda$CDM data at different
redshifts. The filled circles show the two-point correlation functions
measured directly from each simulation output. The mean separation
lengths $\lambda_{\rm part}$ of simulation particles are indicated by
arrows. Dashed lines show the nonlinear prediction by Peacock \& Dodds (1996).
Predictions by the dark halo approach, to which corrections for
the lack of small-scale perturbation power and the finite particle mass
are made (see \S 2), are shown by dotted, long-dashed and solid lines
for the halo-halo term, Poisson term and their sum, respectively.
It is important to keep in mind that the fitting function by Peacock
\& Dodds (1996) may have 10 to 20 percent
uncertainty depending on the cosmological model and scale (Peacock \&
Dodds 1996), thus a small disagreement found
at very small scales in lower redshift cases may be partly due to a
failure of the fitting formula (Jenkins et al.~1998).

At $r> \lambda_{\rm part}$, the halo-halo term almost perfectly
reproduces the simulation results. This reflects the fact that the
evolution of large-scale structure is insensitive on average to the
details of the smaller-scale clustering; incidentally this is why we can
safely apply linear theory on large scales without specifying the
small-scale inhomogeneity.

The evolution of clustering at $r< \lambda_{\rm part}$ is more
interesting.  On these scales, one cannot assign the proper amplitude of
fluctuation spectra predicted from linear theory at $z_{\rm init}$.
Nevertheless at $z=(3-5)$, the almost right amount of power has been
already generated by the nonlinear mode-coupling from the larger scales.
This suggests that the rate of the nonlinear power transfer is
sufficiently rapid.  As a result, at those epochs the small-scale power
in simulations is primarily limited by the finite mass resolution rather
than the lack of the initial fluctuation power on the corresponding
scales.

\subsection{When can we trust the small-scale dark matter clustering 
in simulations ?}

It is remarkable that the two-point correlation function in simulations
on scales below the mean particle separation does not seriously suffer
from the lack of the initial fluctuation power, at least at later
epochs.  Nevertheless the resulting dark matter clustering is still
below the nonlinear prediction due to the finite mass resolution which
cannot be accounted within the 10 to 20 percent uncertainty in the
nonlinear fitting formula (Peacock \& Dodds 1996; Jenkins et al.~1998).
In fact, this systematic introduces error in the computed correlation
function for simulations with lower mass resolution and/or at higher
redshifts.  Then one might naturally ask when the small-scale dark
matter clustering in simulations becomes reliable. We propose an
empirical criterion as follows.

We have already shown that the small-scale dark matter clustering
is primarily limited by the finite mass resolution.  Also the Poisson
term would be dominated by halos which become nonlinear at that
epoch. The characteristic mass for gravitationally collapsed objects,
$M_{\rm NL}(z)$, is approximately given by the condition $\sigma(M_{\rm
NL},z)=1$.  Therefore this mass scale should be much larger than the
simulation particle mass $m_{\rm part}$ so as to reproduce the proper
amplitude of the Poisson term.

Since the small-scale dark matter clustering
mainly comes from the Poisson term (\S 2), the absence of halos with a mass
comparable to and less than the particle mass in a $N$-body simulation
leads to a lack of the clustering amplitude at small scales.
Therefore, in order to reproduce the proper amplitude of the Poisson
term at a certain time, halos with at least a characteristic mass for
gravitationally collapsed objects at that time must be resolved.
Such characteristic mass, $M_{\rm NL}(z)$, is approximately
given by the condition $\sigma(M_{\rm NL},z)=1$, and should be much
larger than the simulation particle mass $m_{\rm part}$ so as to be
resolved in a $N$-body simulation.
This requirement can be translated to the criterion for the critical redshift
$z_{\rm crit}$ when simulations reasonably resolve the dark matter
clustering on scales below the mean separation:
\begin{equation}
 M_{\rm NL}(z_{\rm crit}) = n_{\rm halo} m_{\rm part} .
\end{equation}
In the above we introduce a fudge factor $n_{\rm halo}$ so as to
represent the number of particles that is required to reasonably resolve
a single halo of the nonlinear mass. Usually $n_{\rm halo}$ is typically
supposed to be around 10, but, as far as statistical measures such as
the dark matter two-point correlation functions are concerned, it could
be much smaller; we used $n_{\rm halo} \sim \mu=4$ in plotting Figures
\ref{fig:virgo_scdm} to \ref{fig:hubble_lcdm}, for instance.

\vspace{0.5cm}
\centerline{{\vbox{\epsfxsize=8.7cm\epsfbox{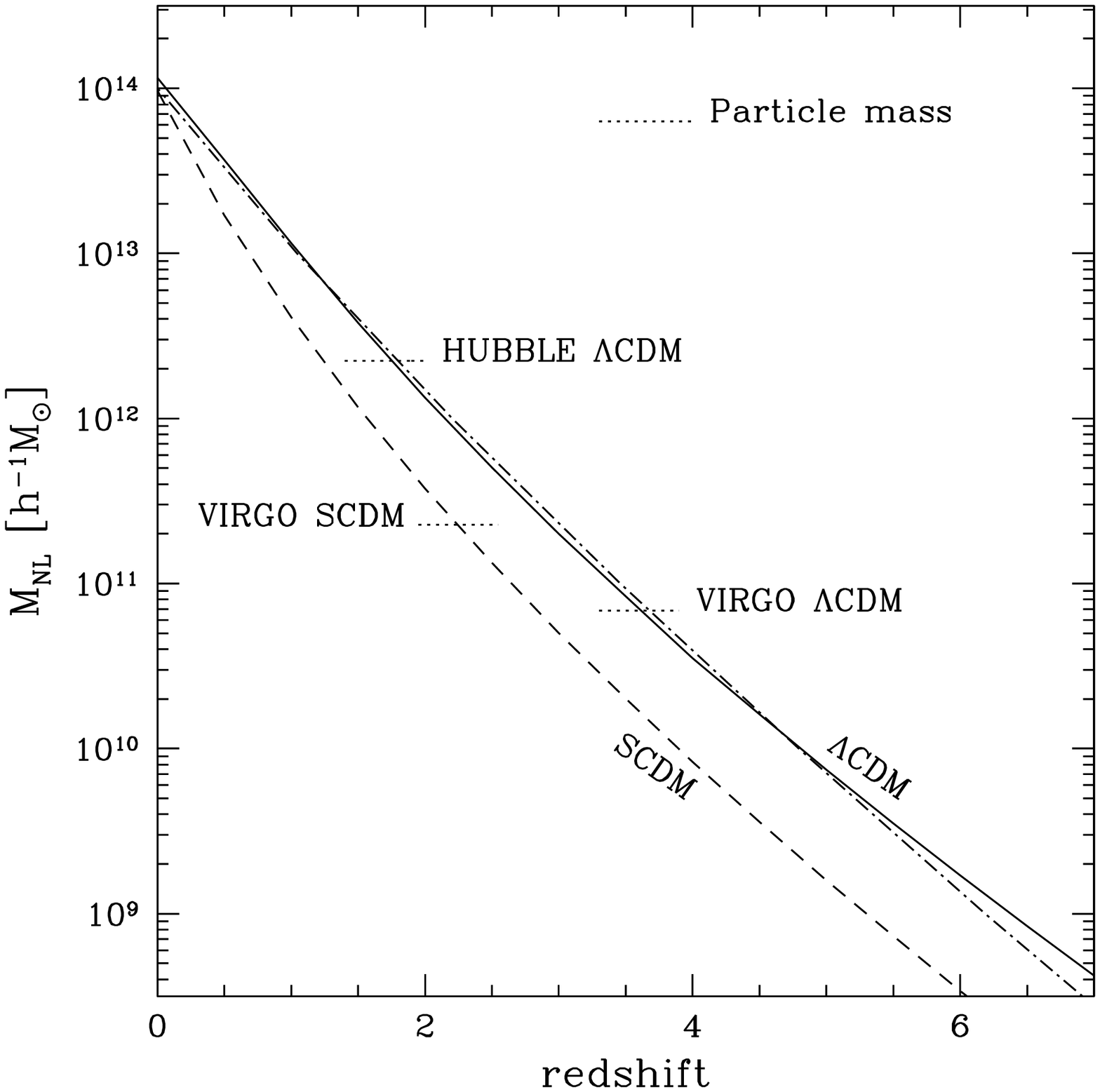}}}}
\figcaption{The nonlinear mass scale $M_{\rm NL}$ defined by
$\sigma_R(M_{\rm NL})=1$ versus redshift. The solid and dashed lines
are for $\Lambda$CDM and SCDM model, respectively. The dotted
horizontal lines shows the particle masses of $N$-body simulations
used in \S 3.
The dot-dashed line shows the fitting function of $M_{\rm NL}$-$z_{\rm
crit}$ relation for $\Lambda$CDM model,
equation  ~(\ref{eq:zcrit-MNL}). \label{fig:sigmar_z.ps}}
\vspace{0.3cm}

In Figure \ref{fig:sigmar_z.ps}, we plot $M_{\rm NL}(z)$ against
redshift for the SCDM (dashed lines) and $\Lambda$CDM (solid lines)
parameters of the present simulations.  The particle masses for the
three simulation models are marked in dotted lines.  The dot-dashed line
shows the approximate relation for the $\Lambda$CDM model that fits the
relation for the $\Lambda$CDM model within 10 percent accuracy for
$0<z<6$;
\begin{equation}
\label{eq:zcrit-MNL}
z_{\rm crit} = 0.72 \left[ \log_{10}\left(\frac{2\times10^{15}h^{-1}M_\odot}
{n_{\rm halo} m_{\rm part}}\right)\right]^{1.25} -1 . 
\end{equation}
As far as the dark matter correlations on scales below the mean particle
separation are concerned, one should not trust the simulation data at
$z>z_{\rm crit}$.  It is important to note that this criterion does {\it
not} mean all results from such $N$-body simulations are unreliable, but
should be applied only to the dark matter correlations at small
scales.

\section{Summary and discussion}

We have examined the reliability of dark matter correlation in
cosmological high-resolution $N$-body simulations on scales below the
mean particle separation.  The particle discreteness effect imposes two
fundamental limitations on those scales; the lack of initial fluctuation
power and finite mass resolution.  We applied dark halo approach to
discuss separately how these two limitations affect the simulation
results at later epochs. We found that nonlinear power transfer below
the mean particle separation (up to the force resolution) is
sufficiently rapid that the reliability of the simulations is primarily
determined by the mass of particles. By a detailed comparison with three
major cosmological simulations, we conclude that at $z>z_{\rm crit}$ the
small scale dark matter correlations in high-resolution $N$-body
simulations are not reliable.  Of course it is impossible to {\it prove}
that the opposite is true at $z<z_{\rm crit}$.  Nevertheless our
comparison indicates also that the small scale dark matter clustering in
a high-resolution $N$-body simulation is reliable down to the
gravitational force resolution length for $z<z_{\rm crit}$ at least as
far as the two-point correlation functions are concerned.  It must be
noted that this criterion does {\it not} mean that all results from such
$N$-body simulations are unreliable.  The limitation of the reliability
of a measurement from $N$-body simulations depends on the quantity one
wants to examine.

The criterion should be applied only to measurements of the dark matter
clustering such as its two-point correlation function and power
spectrum. 
Splinter et al. (1998) concluded that phase sensitive statistics exhibit
strongly the effect of discreteness.  Strictly speaking our conclusion
is not inconsistent with their finding, because the two-point
correlation is insensitive to the phase information and the difference
would largely come from the higher-order correlations that we do not
attempt to compare. In principle our model could be generalized to
higher-order correlations, but there is no reliable empirical model for
those statistics that we can test against, to the quantitative degree of
the Peacock-Dodds prediction for the two-point correlation function.

We believe that the present methodology to test the reliability of the
dark matter clustering in a numerical simulation is original and
powerful, and at least complementary to a more traditional way of
searching for the convergence among different numerical codes with
different resolutions. On the other hand, we have to note that these
approaches are not completely independent; the major ingredients in our
model ``predictions'' including the nonlinear mass power spectrum
(Peacock \& Dodds 1996), the mass function of halos (Sheth \& Tormen
1999), and the density profile of halos (Navarro, Frenk \& White 1996)
are actually {\it calibrated} by high-resolution simulations.  We have
assumed that those calibrations are performed by {\it perfect}
simulations, applied the resulting {\it predictions} to the current
problem, and found that our model remarkably reproduces what we measure
from simulation data.  The present methodology may still suffer from
some calibration uncertainties.
Part of this difficulty is already illustrated in the small but
systematic deviation of our predictions and simulation data on scales
one-order-of magnitude smaller than the mean separation. Most likely
this is due to the gravitational softening effect in simulation
data. Nevertheless it is also true that the model prediction in those
scales may be sensitive to the choice of the inner slope of halo
profile; we adopted the NFW value, 1.0, throughout the paper, but it
could be 1.5 as many recent simulations indicate (Fukushige \& Makino
1997; Moore et al. 1998). Therefore we do not claim that the validity of
the Peacock-Dodds prediction and our halo approach has been tested below
those scales. Our conclusion, however, is valid on scales above $\sim
0.1$ times the mean separation lengths where those details are not
critical.

Exactly for these reasons, we hope to revisit and re-examine the
reliability of the model predictions calibrated by existing simulations
in future.

\acknowledgments
We would like to thank Simon White for valuable comments and his
careful reading of the manuscript, and Gerhard B\"orner for useful
comments.
T.H. thanks Matthias Bartelmann for the hospitality during his
stay at MPA where most of the present work was performed. He also
acknowledges support from Research Fellowships of the Japan Society
for the Promotion of Science.  The simulations used in this work were
carried out by the Virgo Consortium and the data are publicly
available at http://www.mpa-garching.mpg.de/NumCos. 
Numerical computation in the present work was partly carried out at
the Yukawa Institute Computer Facility.
This research was supported in part by the Grant-in-Aid by the
Ministry of Education, Science, Sports and Culture of Japan (07CE2002,
12640231).


\end{document}